\documentclass[11pt]{article}
\usepackage{amsmath, amssymb}
\usepackage{slashed}
\usepackage{verbatim}
\usepackage{latexsym}
\usepackage{euscript}
\usepackage{amsfonts}
\usepackage{verbatim}
\usepackage{cancel}
\usepackage{soul}
\usepackage[normalem]{ulem}
\usepackage[]{array}
\usepackage[]{mathrsfs}
\usepackage[utf8]{inputenc}
\usepackage[T1]{fontenc}
\usepackage{makeidx}
\makeindex
\usepackage{graphicx}
\usepackage{amsmath}
\usepackage{amssymb}
\usepackage{amsxtra}
\usepackage{color}
\def\be{\begin{eqnarray}}
\def\ee{\end{eqnarray}}
\def\0{\nonumber}

\def\tr{{\rm tr}}

\def\sfg{{\sf g}}

\usepackage{slashed}
\usepackage{verbatim}
\usepackage{latexsym}\usepackage{color}
\usepackage{euscript}
\usepackage[curve]{xypic}
\newcommand\EW{\EuScript{W}}

\newcommand\ED{\EuScript{D}}

\def\sfD{{\sf D}}

\def\sfg{{\sf g}}

\normalfont\large
\begin{document}
\vskip 1cm
\begin{flushright}
{SISSA/14/2025/FISI}\\
{\it improved version}
\end{flushright}
\vskip 1cm
\begin{center}

{\LARGE Anomaly footprints in SM+Gravity }

\vskip 1cm

{\large  L.~Bonora$^{a}$\footnote{email:bonora@sissa.it},\\
\textit{${}^{a}$ International School for Advanced Studies (SISSA),\\Via
Bonomea 265, 34136 Trieste, Italy}
}

\end{center}
\vskip2cm
{
{\bf Abstract}. This is a follow-up of \cite{BG24}. A simplified version of the SM plus gravity, put forward there, is presented here and some of its aspects delved into. The basic structure consists of two sectors, left and right, with chirally mirror fermions and scalars, as well as $SU(3)$ and $U(1)$ gauge fields, while the $SU(2)$ gauge fields as well as the metric are in common to both sectors. This structure is dictated by the request to cancel all dangerous anomalies. The left sector consists of the fermion, gauge and scalar fields of the SM, now minimally coupled to gravity. The right sector is a mirror image of the left, with distinct fields, except the metric and the $SU(2)$ gauge potentials. The first new aspect is the proposed and motivated interpretation of the right sector as the dark matter one. The second new subject covered here is Weyl symmetry and its possible application to cosmology and its theoretical fallout on unitarity and renormalization of the model. A background solution of the Weyl invariant theory is derived, which may apply to the very early stages of the universe. This solution also suggests interesting applications to the cosmological constant problem. On the quantum field theory side the subject of Weyl symmetry and Weyl anomalies is reviewed and, among other things, an application of the WZ terms is illustrated  to the problem of one-loop quantization of the model which may avoid negative norm states.

\section{Introduction}

Gauge and gravitational anomalies in local field theories come in two species. In \cite{BG24} these two types were labeled type O (obstructive) and type NO (non-obstructive). The reason of the partition is due to their drastically different nature. The former are a symptom that  (chiral fermion) propagators do not exist. For they signal topological obstructions to invert the corresponding Dirac operators, made precise by the family's index theorem of Atiyah and Singer\footnote{A full treatment of O-type anomalies can be found in \cite{I}.}; when present in a theory, they make impossible its quantization. The NO-type anomalies are simply quantum effect, they signal that a symmetry is violated at the quantum level but have no obstructive effect on propagators and do not endanger quantization. Needless to say chiral theories that contain O-type anomalies must be discarded. The minimal standard model (MSM) is free of the latter. But when we couple it to gravity new anomalies come into play. They mostly cancel out due to the particular algebraic combination of the fields in it. But some residual odd-parity trace anomalies suvive. They are generated by their coupling to gravity and are built with the $SU(2)$ gauge fields.

In \cite{BG24} a theory was tentatively proposed which is free of all O-type anomalies. It consists of mirror right-handed multiplets of fermions added to the left-handed multiplets that define the MSM.
Altogether it features a sort of two distinct standard models coupled to gravity, which have in common the $SU(2)$ gauge fields, while they have distinct $SU(3)$ and $U(1)$ gauge sectors, as well as two distinct gravity sectors, each one with its own metric, and distinct scalar sectors. It was also shown that this model can be extended with the addition of dilaton fields that guarantees Weyl invariance. This model is type-O anomaly free and has a quality, it answers the question: why does nature use particles and antiparticles of only one handedness to build up our universe while disdaining particles and antiparticles with the other? Having to include the latter in the cast is an anomaly footprint. 

But once its merits are recognized, how do we interpret the particles that mirror the MSM (the right sector)? One possibility is to use the modeler's privilege and simply imagine a mechanism that assign very large masses and exclude them from any realistic possibility of experimental detection, as was done in part in the last-but-one section of \cite{BG24}. A more appealing possibility, not altogether disconnected from the previous one, is to interpret them as the dark matter sector. We shall discuss this point later on. But before that we should clarify the nature of the above outlined  theory. It is an attempt to assemble a theory starting from different (and non-homogeneous) components, the MSM on one side and general relativity (GR) on the other. They are not homogeneous in the sense that while the MSM is a unitary and renormalizable quantum field theory, GR is a classical theory yet without any known quantum field theory UV completion. In a theory describing the four fundamental forces, there is a stage where the SM, a quantum field theory, live together with GR.  This is the stage where the scale of energy is such that life is possible. But it is natural to ask ourselves what happens when the reference energy increases and we trace back the evolution of the universe toward the big-bang beginning with a constantly increasing scale of energy.

The most natural thing to do is to couple the SM fields to gravity in the simplest way, the minimal one, and deal with the resulting theory as a unique theory and quantize it. As a first step in quantization one has to deal with the anomalies. This is what was done in \cite{BG24}, and we reconsider here.  The theory complies with the first elementary requirement, absence of O-type anomalies. It contains only renormalizable terms, that is it is a power-counting renormalizable theory, although this is not enough to secure renormalization and unitarity\footnote{This is due to the presence of zero dimension fields, like the metric, which allow for infinite many interaction terms in the action, \cite{piguetsorella}}.  On the other hand expecting a unitary and renormalizable field theory with a limited number of fields including gravity  would seem to be overly naive. 
In other words the theory put forward in \cite{BG24} and studied here is at an initial stage, but this does not mean that we cannot extract from it useful and non-trivial information, such as the doubling of dofs with opposite chiralities required by the anomaly cancelation. This bottom-up approach is not of course the only possible one. There are several top-down approaches, notably the ones based on supergravity field theories or on superstring theories. They rely on the effective field theory (EFT) techniques \cite{Burgess}, which goes essentially back to the Wilsonian philosophy. It stems from the idea of integrating out the high energy (momenta and masses, high with respect to a scale parameter) degrees of freedom and retain only the renormalizable terms among those obtained, relying on the expectation that the higher order terms should be suppressed by the inverse power of the scale factor. Although obtained through a different path the result in \cite{BG24} is similar. For this reason that action is to be viewd in the spirit of the EFT approach.

Another aspect analyzed in \cite{BG24} as well as below is Weyl (conformal) symmetry. It is often stated in the literature that at very high energy masses and other dimensionful constants become irrelevant. Introducing a dilaton field it is possible to reformulate field theories containing dimensionful constants in a Weyl invariant way, whereby masses and dimensionful constants need not be naively set to zero in a high energy regime, but appear in the action in such a way that their values are irrelevant. This symmetry however is challenged by trace anomalies. The latter are NO-type and do not endanger quantization, but only break conformal symmetry. If our aim is to preserve conformal invariance,  there is a way to restore it by means of Wess-Zumino terms, which require introducing an extra field. This amounts to adding them to the action by identifying the extra field with the dilaton. These additional terms are another anomaly footprint. 

The purpose of this conference paper is to present a simplified version of the theory put  forward in \cite{BG24}, in particular only one metric, instead of two, is considered, and then add a few, although still largely incomplete, remarks on two particular aspects:  the interpretation of the mirror model and conformal invariance. The paper is organized as follows. Section two is a presentation of the theory introduced in \cite{BG24} in a simplified form. Section 3 is devoted to a comment on the  physical interpretation of the right sector of the theory. In section 4 Weyl invariance is presented and discussed, together with a possible application to the description of physics of the early universe. Section 5 is dedicated to the trace anomalies and their cancelation by means of WZ terms, together with their possible implications for unitarity.

\section{A left-right symmetric model}

The following model was introduced in \cite{BG24} with two metrics. Here we consider a simplified version in which there is only one metric. The fermion matter part is based on the same multiplet as the MSM with the addition of a right-handed sterile neutrino. In the usual SM notation it is 
\be
\begin{matrix} {\sf G}/fields & \quad SU(3)\quad &\quad SU(2)\quad &\quad U(1)\quad\\
\left( \begin{matrix} u \\ d\end{matrix} \right)_{\! L} & 3&2&\frac 16\\
{(u_R)^c} &  \bar 3 &1&-\frac 23\\
{(d_R)^c} & \bar 3 &1&\frac 13\\
\left( \begin{matrix} \nu_e \\ e\end{matrix} \right)_{\! L} & 1&2&-\frac 12\\
{(e_R)^c} &  1 &1&1\\
{(\nu_R)^c} &  1 &1&0
\end{matrix}\label{Lspectrum}
\ee
where $ X^c$ represents the Lorentz conjugate spinor of $X$, i.e. $X^c=\gamma_0 C X^\ast$. This multiplet couples to a gravitational metric and connection, and to the $SU(3)_L\times SU(2)\times U(1)_L$ gauge fields. In \cite{BG24} it was shown that all the anomalies cancel out except for  4 units of the trace anomaly with density $F\ast F$, due to the gauge field   $F\equiv F^{\mathfrak su(2)}$,  computed in the doublet representation of $\mathfrak su(2)$. 

In the multiplet \eqref{Lspectrum} the doublet $\left(\begin{matrix} u\\ d\end{matrix}\right)_L$  describes left-handed particles and right-handed antiparticles, while the singlets $u_R$ and $d_R$ represent right-handed particles and left-handed antiparticles, and similarly for the leptons.

The main difference with the MSM is that the spectrum is completed by a right-handed multiplet
\be
\begin{matrix} {\sf G}/fields & \quad SU(3)\quad &\quad SU(2)\quad &\quad U(1)\quad\\
\left( \begin{matrix} u' \\ d'\end{matrix} \right)_{\! R} & 3&2&\frac 16\\
{(u'_L)^c} &  \bar 3 &1&-\frac 23\\
{(d'_L)^c} & \bar 3 &1&\frac 13\\
\left( \begin{matrix} \nu'_e \\ e'\end{matrix} \right)_{\! R} & 1&2&-\frac 12\\
{(e'_L)^c} &  1 &1&1\\
{(\nu'_L)^c} &  1 &1&0
\end{matrix}\label{Rspectrum}
\ee
coupled to the gravitational metric and connection. It also couples to the $SU(3)_R\times SU(2)\times U(1)_R$ gauge fields. The anomaly analysis of this mirror multiplet\footnote{A mirror sector of the SM has been considered earlier in the literature, see \cite{hodges93,mohapatra97,bai14}, with various purposes related to cosmology.}  is the same as for the left-handed one except for the sign of the trace anomaly due to the gauge field   $F\equiv F^{\mathfrak su(2)}$,  which is opposite. Therefore the overall sum of the anomalies of the system vanishes.

In the multiplet \eqref{Rspectrum} the doublet $\left(\begin{matrix} u'\\ d'\end{matrix}\right)_R$  describes right-handed particles and left-handed antiparticles, while the singlets $u'_L$ and $d'_L$ represent left-handed particles and right-handed antiparticles, and similarly for the leptons. 

Of course we should write three families of left-handed and three families of right-handed fermions. But since the physics that intetwines different families will not be discussed here, one single family will do.

We shall call these two intertwined theories, with field content \eqref{Lspectrum} and \eqref{Rspectrum}, ${\cal T}_L$ and ${\cal T}_R$, respectively. The overall theory is free of type O anomalies. We denote it simply by ${\cal T}={\cal T}_L \cup {\cal T}_R$. The symbol $\cup$ is because the two half theories have the metric and the SU(2) gauge potentials in common.

{\bf Important.} Both multiplets couple to the same $SU(2)$  gauge fields. Only in this case do all anomalies cancel! We remark that, since, contrary to \cite{BG24} there is only one metric, the presence of the sterile neutrinos
$\nu_R$ and $\nu'_L$ is not necessary in order to cancel all type O anomalies.
\vskip 0.3cm
Let us see explicitly in the sequel the various possible pieces of the relevant actions. Let us start with the fermion kinetic actions. We have 
\be
S_f^{(+)}\equiv S_{fR}
&=&\int d^4 x \, \left(\sqrt{g}\, i\overline {{\psi'}_{R}}  
\gamma^a
 e_a^{\mu}
\left(\ED_\mu^{(+)}+\frac 12 \omega_\mu \right)\psi'_{R}\right)(\widehat x)\label{fermionaction+}
\ee}
where $\psi'_R$ represents the right-handed multiplet \eqref{Rspectrum}, and 
\be
\ED^{(+)}_\mu=\partial_\mu +{\sfg}_X^+ X^{(+)}_\mu +{\sfg}_WW_\mu +{\sfg}_B^+B_\mu^{(+)} \label{XWQ+}
\ee
As usual $ \omega_\mu=\omega_\mu^{ab}\Sigma_{ab}$ represents the spin connection corresponding to the metric $g$ and $\Sigma_{ab}$ the anti-hermitean Lorentz generators. For the left sector 
\be
S_f^{(-)}\equiv S_{fL}
&=&\int d^4 x \, \left(\sqrt{g}\, i\overline {{\psi}_L}  
\gamma^a
 e_a^{\mu}
\left(\ED^{(-)}_\mu+\frac 12 \omega_\mu\right)\psi_{L}\right)(\widehat x)\label{fermionaction-}
\ee
where $\psi_L$ represents the left-handed multiplet \eqref{Lspectrum}, and
\be
\ED^{(-)}_\mu=\partial_\mu + X^{(-)}_\mu +W_\mu +B_\mu^{(-)}
\ee
The symbols $X^{(\pm)}_\mu,W_\mu,B^{(\pm)}_\mu$ refer to the $SU(3)_{R/L}, SU(2)$ and $U(1)_{R/L}$ potentials, respectively.  Of course each potential has its own distinct coupling to the fermions, which can be made explicit through a redefinition of the potentials.

Let us recall that the symbol such as $(\psi_R)^c$ (for instance $(u_R)^c, (d_R)^c, ...$) can be rewritten as
\be
 (\psi_R)^c= \gamma^0 C \psi_R^*= \gamma^0 C P_R^* \psi^*=P_L \gamma^0 C\psi^*=P_L \psi^c = (\psi^c)_L.\label{psiLpsiR}
\ee
Inserted into the kinetic term, this gives
\be
\int d^4x\sqrt{g}\, \overline {(\psi^c)_L} \,\gamma^\mu(\partial_\mu +\frac 12 \omega_\mu )(\psi^c)_L=\int d^4x\sqrt{g}\, \overline {\psi_R}\, \gamma^\mu(\partial_\mu +\frac 12 \omega_\mu )\psi_R\label{LcR}
\ee
taking account that $\Sigma_{ab}$ are anti-hermitean, using an overall transposition and a partial integration. Therefore the kinetic term of the multiplet \eqref{Lspectrum}, coupled only to the metric, splits into 16 independent Weyl fermion kinetic terms, 8 left-handed and 8 right-handed, with opposite contribution to the odd parity trace anomaly. 

The SU(2) gauge field action has the usual form
\be
S^{\tiny{SU(2)}}_g =- \frac 1{4{\sf g}^2}  \int d^4 x \,\sqrt{g} \, \tr \left( g^{\mu\mu'} g ^{\nu\nu'}  F_{\mu\nu } F_{\mu'\nu'} \right)\label{actiongf+-}
\ee
where $F_{\mu\nu }=  d V +\frac 12[ V, V]$ is the curvature of the SU(2) gauge field\footnote{In \cite{BG24} two $SU(2)$ gauge couplings were introduced, one for each sector; however the cancelation of $SU(2)$ gauge-induced odd trace anomalies requires that there be only one coupling.}.

For the groups $ SU(3)_L\times SU(3)_R$ and $ U(1)_L\times U(1)_R$ we have instead
$S_{g}^{(+)}+ S_{g}^{(-)}$ with
\be
S_{g}^{(\pm)} =- \frac 1{4 \sfg_{\pm}^2} \int d^4 x \,\sqrt{g} \, \tr \left( g^{\mu\mu'} g^{\nu\nu'}  F^{(\pm)}_{\mu\nu } F^{(\pm)}_{\mu'\nu'} \right)\label{actiongfae+-2}
\ee
where $ F^{(\pm)}_{\mu\nu }=  d V^{(\pm)} +\frac 12[ V^{(\pm)}, V^{(\pm)}]$ and  $F^{\pm}$ denotes the curvatures of the $SU(3)_R$ and $ U(1)_R$,  and $SU(3)_L$ and $U(1)_L$ potentials, respectively. $S_g^{(\pm)}$ is supposed to represent the sum of both for $SU(3)$ and $U(1)$ gauge action with distinct couplings, which can be absorbed, as usual, in a redefinition of the respective gauge potentials.

The action for the metric is the usual EH action with different cosmological constants  in the left and right sector 
\be
S^{(\pm)}_{EH}=- \frac 1{2\kappa}   \int d^4 x \,\sqrt{g} \left( R+ {\mathfrak c}_\pm\right)\label{EH+-}
\ee
Here $R$ is the Ricci scalar, $\kappa$ the  gravitational constant and ${\mathfrak c}_\pm$ the left/right cosmological constant. 

In the MSM we need also a couple $H_\pm$ of complex scalar fields, which minimally couple to the metric $ g_{\mu\nu}$ and are a doublet under $SU(2)$. The corresponding actions in the two sectors are given by
\be
S_{d}^{(\pm)}= \int d^4\widehat x \,\sqrt{g} \left[ g^{\mu\nu} \ED_\mu H_\pm^\dagger \ED_\nu H_\pm -  M_\pm^2 H_\pm^\dagger H_\pm -\frac {\lambda_\pm}4 \left(H_\pm^\dagger H_\pm\right)^2\right]\label{Hscalar+-}
\ee
where $\ED_\mu = \partial_\mu -i\sfg W_\mu$, and $W_\mu$ is the $SU(2)$ gauge field.

So far we have considered pieces of action representing matter minimally coupled to the metric and to gauge potentials. Now we need the interaction among matter fields. This is given by the Yukawa couplings. They split into left and right parts. For instance, for SU(2) doublets we have
\be
S_{YdL} = \frac {y^-_{H_d}}2 \int d^4 x \,\sqrt{g}\left( \overline{\psi_{dL}}\,{H}_{d-} \chi_{sR}\right)+ h.c.\label{yukawadL+conj}
\ee
where $\psi_{dL}$ is a left-handed SU(2) doublet, $  H_{d-}$ is also an $SU(2)$ doublet, conjugate to the $\psi_{dL}$ one in the inner product of the $SU(2)$ doublet representation space, while $\chi_{sR}$ is a right-handed singlet, all of them belonging to ${\cal T}_L$. Similarly, for ${\cal T}_R$,
\be
S_{YdR} = \frac {y^+_{H_d}}2 \int d^4\widehat x \,\sqrt{g}\left( \overline{\psi'_{dR}}\,{H}_{d+} \chi'_{sL}\right)+ h.c. \label{yukawadR+conj}
\ee

Let us write $S_f= S_f^{(+)} +  S_f^{(-)} $, $S_g=S^{SU(2)}_g+ S_g^{(+)} +  S_g^{(-)} $, $S_{EH}= S_{EH}^{(+)} +  S_{EH}^{(-)} $, $S_d= S_d^{(+)} +  S_d^{(-)} $ and $S_Y= S_{YdL} +  S_{YdR} $. Then for the total action of  our model minimally coupled to gravity we can tentatively set
\be
S=S_f+S_g+S_{EH} +S_{d} + S_Y \label{totalaction}
\ee
This theory is invariant under $SU(2)$, as well as  $ SU(3)_L\times SU(3)_R$ and $ U(1)_L\times U(1)_R$, gauge transformations. It is also invariant under diffeomorphisms and local Lorentz transformations. Concerning the discrete symmetries, each term of the sum \eqref{totalaction} is $CP$ and $T$ invariant in the left and right sector separately; but in general it is not $P$ and $C$ invariant. $P$ invariance of the overall $S$ requires that all constants and masses appearing in $S$ with labels + and - be equal, i.e. ${\mathfrak c}_-= {\mathfrak c}_+$, etc. Moreover both left and right parts separately have the same symmetries. We say that $\cal T$ is left-right or chirally symmetric. For conciseness, we shall call left the matter fields of ${\cal T}_L$ and right the matter fields of ${\cal T}_R$, of course with the exclusion of the metric and the $SU(2)$ gauge fields.

Eq.\eqref{totalaction} is likely to be the minimal form of the anomaly-free action including both  MSM and gravity. As was mentioned before and shown in \cite{BG24}, both ${\cal T}_L$ and ${\cal T}_R$ are separately free of O-type anomalies, except for the trace anomaly whose density is $\sim F * F$. Putting together the two halves has the effect of canceling also this anomaly. In this statement there is no claim of completeness and up to here we do not consider the problem of renormalization and unitarity. $S$ is obtained by putting together the indispensable ingredients. It complies however with the first essential requirement for an effective theory: it is free of obstructive anomalies, so that all the propagators and vertices are well defined and a perturbative quantization can be carried out. The next condition for effectiveness is unitarity. If, in addition, the theory happens to be renormalizable, then it is UV complete.

\section{Dark matter?}

The theory described by $S$ splits into two halves, each with distinct scalar and fermion matter components. Also the gauge groups $ SU(3)_L\times SU(3)_R$ and $ U(1)_L\times U(1)_R$, respectively, are distinct, while the metric and the $SU(2)$ gauge fields are the same on both sides. The left matter and the right matter interact only via the latter fields and in no other way. For instance the left fermions (left-handed particles and right-handed antiparticles) interact among themselves strongly via the $ SU(3)_L$ gauge bosons and electromagnetically via the $ U(1)_L$ potential. Thanks to the Yukawa couplings they interact with the left doublet of scalars. They interact also weakly via the $SU(2)$ gauge fields and gravitationally via the metric. However with the mediation of the latter they interact also with the right fermions (right-handed particles and left-handed antiparticles). An example of these types of interaction is the scattering of a left fermion with a right one via the exchange of an $SU(2)$ gauge boson or a graviton as in figure 1 below.  In an analogous way the left scalar fields interact among themselves via the scalar potential, then they interact with the metric and the $SU(2)$ gauge fields, and, via the latter, with the right scalars.
\begin{figure}[h]
    \centering
   \includegraphics[scale=0.6]{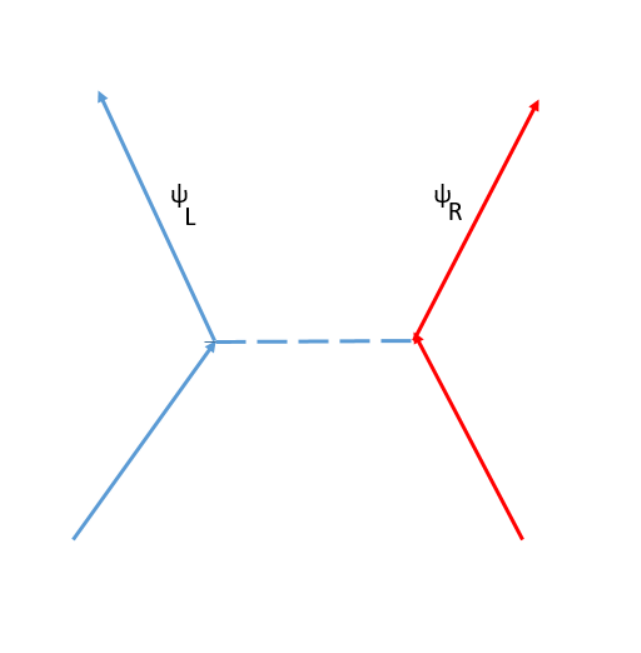}
 \caption{\emph{\small A scattering between left and right fermions mediated by $SU(2)$ gauge bosons or gravitons (dashed line).}}
   \label{fig:AA}
\end{figure}

For the right-handed fermions and scalars we have of course a mirror description. If we imagine that this is the theory at the basis of the evolution of the universe after what is called the grand-unification era, $10^{-35}sec$ after the beginning, there is certainly something missing, at least one or more fields (which, for simplicity, are not explicitly written down) that can describe the inflationary period and possibly a quintessence field, if the cosmological constants are not enough to describe the present accelerating expansion. But all the rest is there. In particular the left-handed part with the addition of an inflaton field, for instance, can effectively describe, resting on the background of a LFRW metric, the physics of the universe's evolution (inflation, reheating, particle creation and density perturbations, perhaps even dark energy, for reviews see \cite{parker69,parker71,ford77,traschen90,Mukhanov92,shtanov95,kofman95,Greene97,Brandenberger03,Tsuchikawa03,Vasquez21,Ford21,bambi2021,Kallosh25} and references therein). Then the question is: if this is the correct picture, what does the right part represent in it? It does not take much imagination to see in it a candidate for the dark matter. This part (the right one) of the total matter+energy evolves in a way parallel to the ordinary matter+energy, although with different coupling constants, masses and cosmological constant. And at the present time energy scale the interaction between the two is limited to the gravitational one, since the common weak force between left and right matter has a too short range to be effective (but of course this is not the case in very high energy scattering phenomena that can involve left and right matter). What makes the difference between the two half theories are the couplings, scalar masses and cosmological constant, beside the handedness of the respective fermions (representing at the massless level, for instance, left-handed particles and right-handed antiparticles, on the left side, and right-handed particle and left-handed antiparticles on the right one). For instance, the evolution in the right part (supposedly the dark matter) might be sensibly different from the ordinary matter in what concerns inflation, particle creation, density perturbations and so on. Certainly we cannot `see' the right-handed world, but at most `feel' it via gravity, and also the weak force if the energy is high enough. This is suggestive, but it remains for us to explain many aspects, of which the most important is  why the amount of dark matter is more than five times larger than the visible matter. And figure out experiments that permit to falsify the idea. 

This said, can we say on this subject something more  than the above rather generic considerations? The answer, somewhat surprisingly, is yes. The literature on dark matter is vast, with a variety of different ideas and proposals. It is classified in cold, warm and hot dark matter. Focusing on the most popular cold dark matter (CDM), it may be made of baryonic objects, i.e. made of the SM baryons, like MACHOs (massive astrophysical compact halo objects) or primordial black holes. The non-baryonic matter might be made of particles: massive neutrinos, axions, or weakly interacting particles present in supersymmetric models: neutralinos (a mixture of supersymmetric partners), fotinos, Binos,... popularly denoted by the acronim WIMPs  (weakly interacting massive particles). WIMPs have been considered as among the most likely candidates for dark matter. They are massive, thus they feel gravity. They belong to the same left sector as the SM particles, and usually are taken from supersymmetric extensions of the SM, or supergravity, or superstring inspired models. But the essential aspect is that they weakly interact with the SM sector. In this sense the right sector, ${\cal T}_R$, can well play the same role and one can simply parasitize, at least in a phase of research designing, the WIMP literature, see \cite{bergstrom00,bertone05,arina12,drees12,bambi2021,arbey21} for reviews. In this sense the first striking aspect of it is the so-called `WIMP miracle': with the freeze-out mechanism, WIMPs achieve the relic density for dark matter appropriate to reproduce the latest experimental data, $\Omega_{dm}\approx 0.25$. Although it is not clear what form the mirror matter will take in this new description, whether macroscopic bodies, gas of neutral particles, like the neutrinos in the left sector, or neutral atoms, or all of them together. This depends very much on the details of the evolution of the mirror world. 

No doubt the idea is suggestive and not airy-fairy. It has also the virtue of unveiling a mistery: why does nature use only particles and antiparticles of one handedness to build up our universe while disdaining  particles and  antiparticles of the opposite one? This should not be confused with the problem of baryon asymmetry, which has to do with the fact that our universe is made essentially by matter (as opposed to antimatter), but it is rather an additional puzzle. It is pleasing that several papers in the literature have considered that baryon asymmetry and its solution (baryogenesis) might be interrelated with dark matter, \cite{boucenna,racker14}. This is an additional motivation to study the two puzzles (baryon asymmetry and left-right asymmetry) in the framework of $\cal T$.

\section{Weyl symmetry}

It is a common belief that whatever local theory we consider, when the energy regime grows very large, masses and dimensionful constants become irrelevant. A related point of view is that these constants may not be true constants, but vacuum expectation values of suitable fields that condense at low enough energies; so that in a fundamental theory only dimensionless constants and fields will appear. In any case, since in any such theory there is no explicit scale, we expect it to be invariant under a rescaling of the metric. Under general conditions this means that the theory is invariant not only under rigid rescalings of the metric, a property referred to as scale symmetry, but also under local ones, which defines conformal or Weyl symmetry. The full $\cal T$ theory is not invariant under local Weyl transformations 
\be
g_{\mu\nu} \rightarrow  e^{2\omega} g_{\mu\nu}\label{Weyltransf}
\ee
where $\omega $ is a local parameter, but several pieces thereof are. They are $S_f, S_g$ and $S_Y$. The remaining pieces are not in general Weyl invariant as they contain dimensionful constants. But it is actually very simple to transform a local theory into a Weyl invariant one by adding a new field, $\varphi$, the dilaton. Under the same Weyl rescaling it transforms as $\varphi \rightarrow \varphi+\omega$. The procedure is as follows. Let us start from the Christoffel symbols.
They transform as
\be
\Gamma_{\mu\nu}^\lambda \to \Gamma_{\mu\nu}^\lambda + \delta_\mu^\lambda\, \partial_\nu \omega +  \delta_\nu^\lambda\, \partial_\mu \omega- g_{\mu\nu} g^{\lambda\rho} \partial_\rho\omega\label{Christtransf}
\ee
We can construct Weyl-invariant Christoffel symbols as follows
\be
\widetilde \Gamma_{\mu\nu}^{\lambda} = \Gamma_{\mu\nu}^\lambda - \left(\delta_\mu^\lambda\, \partial_\nu \varphi +  \delta_\nu^\lambda\, \partial_\mu \varphi - g_{\mu\nu} g^{\lambda\rho} \partial_\rho\varphi\right)\label{Christ1}
\ee
We can use these symbols to build the Riemann and Ricci tensor. The latter is
\be
\widetilde R_{\mu\nu} = R_{\mu\nu} +2\partial_\mu\partial_\nu \varphi+g_{\mu\nu}\square\varphi +2 \partial_\mu \varphi \partial_\nu \varphi-2 g_{\mu\nu} \partial \varphi \cdot \partial\varphi\label{tildeRiccimunu}
\ee
and Ricci scalar is
\be
\widetilde R = R + 6\left(\square\varphi -\partial \varphi\cdot \partial\varphi\right), \label{Rscalar}
\ee
$\widetilde R_{\mu\nu}$ is Weyl invariant, while $\widetilde R \rightarrow e^{-2\omega} \widetilde R$. For the sequel let us remark that if we write $\widetilde g_{\mu\nu} = e^{2\varphi} g_{\mu\nu}$ we can write
\be
\widetilde R_{\mu\nu} (g) = R_{\mu\nu} (e^{2\varphi}g)\label{basic}
\ee
where the entry $g$ in the round brackets is a shorthand for the metri $g_{\mu\nu}$.

 Now the recipe is as follows. In the action we replace $R$ with $ \widetilde R$. Then we multiply every dimensionful constant of mass dimension $s$ by the factor $e^{-s\varphi}$. When applied to scalar fields we replace the simple derivatives  $\partial_\mu$ by:
\be
{\sfD}_\mu = \partial_\mu +\partial_\mu \varphi
\ee
The pieces  $S_f, S_g$ and $S_Y$ need not be modified because they are already Weyl invariant. In the sequel we introduce two distinct dilatons $\varphi_\pm$, one for each sector. They behave exactly as the just introduced $\varphi$.

Specifically, for $\cal T$ we have the following modifications. The EH part becomes
\be
S^{(c\pm)}_{EH} =-\frac 1{2\kappa} \int d^4x \sqrt{g}\, e^{-2\varphi_\pm} \left(\widetilde R_\pm + {\mathfrak c}_\pm\, e^{-2\,\varphi_\pm}\right)\label{ScEH}
\ee
where $ \widetilde R_\pm = R + 6\left(\square\varphi_\pm -\partial \varphi_\pm\cdot \partial\varphi_\pm\right)$, 
and the doublet scalar action becomes
\be
S_{d}^{(c\pm)}= \int d^4 x \,\sqrt{g} \left[ g^{\mu\nu} \ED^{\pm}_\mu H_\pm^\dagger \ED^{\pm}_\nu H_\pm -e^{-2\varphi_\pm}  M_\pm^2 H_\pm^\dagger H_\pm -\frac {\lambda_\pm}4 \left(H_\pm^\dagger H_\pm\right)^2\right]\label{Hscalar+-}
\ee
where $\ED^{\pm}_\mu= \partial_\mu + \partial_\mu \varphi_\pm- i\sfg W_\mu=\sfD_\mu- i\sfg W_\mu$ , $W_\mu$ being a gauge field valued in the $SU(2)$ Lie algebra representation  to which $H_\pm$ belongs.

The Weyl invariant generalization of $\cal T$ is therefore
\be
S^{(c)}= S_f+S_g+S_Y + S_{EH}^{(c)} + S_d^{(c)}\label{totalactionc}
\ee
where $ S_{EH}^{(c)}=  S_{EH}^{(c+)}+  S_{EH}^{(c-)}$  and $ S_d^{(c)}= S_d^{(c+)}+ S_d^{(c-)}$.

For later discussion we add also Weyl invariant action terms. One is the higher derivative term $S_C= S_C^{(+)}+ S_C^{(-)}$ where
\be
S _C = \frac 1{\eta}  \int d^4x \sqrt{g}\, C_{\mu\nu\lambda\rho} C^{\mu\nu\lambda\rho}\label{Weyltensoraction}
\ee
$C_{\mu\nu\lambda\rho}$ is the Weyl tensor (invariant under Weyl transformations). If we  disregard total derivatives in the action, \eqref{Weyltensoraction} can be replaced by
\be
S'_C =  -\frac 2\eta \int d^4x \sqrt{g} \left( - R_{\mu\nu} R^{\mu\nu}+ \frac 13 R^2\right)  \label{Sc'W}
\ee 
The quadratic terms in brackets contain higher derivative kinetic and interaction terms.

Another Weyl invariant action can be constructed for a scalar field $\Phi$.	
\be
S_\Phi = \frac 12 \int d^4 x \,\sqrt{g} \left(\partial_\mu \Phi \partial^\mu \Phi +\frac 16 R\, \Phi ^2\right) \label{scalaraction}
\ee
where $R$ is the Ricci scalar. 

In the literature the action $S^{(c\pm)}_{EH} $ is sometimes modified with the addition of a non-minimal gravitational coupling, so that it becomes:
\be
S^{(c'\pm)}_{EH} =-\frac 1{2\kappa} \int d^4x \sqrt{g}\, \left(e^{-2\varphi_\pm}+ \zeta_{h\pm}H_\pm^\dagger H_\pm \right)\left(\widetilde R + {\mathfrak c}_\pm\, e^{-2\,\varphi_\pm}\right)\label{ScEH}
\ee
where $\zeta_{h\pm}$ are dimensionless couplings

The theory defined by \eqref{totalactionc}, with the possible addition of \eqref{Weyltensoraction}, and $S^{(c'\pm)}_{EH}$ instead of $S^{(c\pm)}_{EH}$, 
 has the same symmetries as $S$, \eqref{totalaction}. In particular it is invariant under the diffeomorphisms spanned by the parameter $\xi^\mu$, with the dilaton transforming as
\be
\delta \varphi_\pm= \xi ^\mu \partial_\mu \varphi_\pm\label{varphidiff}
\ee
In addition it is conformally invariant. It should be duly appreciated that conformal invariance of the action $S^{(c)}$ precisely embodies the idea that at high energies constants and masses are indefinite. For instance the mass factor $ M_\pm e^{-2\varphi_\pm}$, and other similar factors, can take any value, from 0 to $\infty$, without changing the value of the action. We shall call the new theory $\cal TW$.
\vskip 0.2cm
Refs.:\cite{ghilencea21,ghilencea23,ghilencea24,mohammedi2024,roumelioti24}, see also the reviews \cite{scholz2018,rachwal2022}.

\subsection{Meaning and import of conformal invariance}

The theory outlined in the previous section is classical, its quantum aspects will be considered later. But suppose that one such theory is adherent to the physics of fundamental interactions in a certain range of energy and a semiclassical approach makes sense, we face a problem of interpretation: what is the significance of conformal invariance? Although a local symmetry, conformal invariance has characteristics that differentiate it from ordinary gauge theories. To start with, the `gauge field', in the version presented here, is a scalar. In this section we wish to treat it as a classical field and treat conformal symmetry as an ordinary rigid symmetry, like $O(N)$, for instance, in models with the same name. This means  that it is a physical symmetry. Different configurations of $\varphi$ are physically distinct, although with the peculiarity that their description differ by a symmetry operation. Different configurations of $\varphi$ define equivalent solutions of $S^{(c)}$, but considering them from an energy regime where conformal invariance is not anymore a symmetry, they may describe a very different physics. 

As a first step let us plug our conformal invariant theory in a cosmological framework. To this end we search for classical solutions of time-dependent, but space-independent, fields. and, to be concrete, we choose for the metric the Friedmann-Robertson one:
\be
ds^2= dt^2 -a^2(t) \left( \frac {dr^2}{1-kr^2} +r^2 d\theta^2 +r^2 \sin^2\theta\, d\phi^2\right) \label{FRmetric}
\ee
Let us recall that with this metric we have
\be
&&g_{tt}=g^t_t=1, \quad\quad g^r_r=g^\theta_\theta=g^\phi_\phi=1\0\\
&& R_{tt}=R^t_t = -3 \frac {\ddot a}a ,\quad\quad R^r_r=R^\phi_\phi=R^\theta_\theta= - \left( \frac {\ddot a}a +2 \frac {{\dot a}^2}{a^2} +2 \frac k{a^2}\right)\label{RFRW} \\
&& R =  -6 \left( \frac {\ddot a}a + \frac {{\dot a}^2}{a^2} + \frac k{a^2}\right)\0
\ee
As usual, dots denote derivative with respect to time.
For simplicity let us start with a conformal invariant action for a metric, a dilaton and a scalar field $\Phi$:
\be
S_1^{(c)}=-\frac 1{2\kappa} \int d^4x \sqrt{g}\, e^{-2\varphi} \left(\widetilde R + {\mathfrak c}_\pm\, e^{-2\,\varphi}\right)+ \frac 12  \int d^4 x \,\sqrt{g} \left[ g^{\mu\nu} \sfD_\mu \Phi \sfD_\nu \Phi -e^{-2\varphi}  m^2 \Phi^2 -\frac {\lambda}4\Phi^4\right]\label{Strunc}
\ee
where $\sfD_\mu = \partial_\mu+ \partial_\mu \varphi$ and the $\zeta$ non-minimal coupling has been dropped. This is clearly a  drastically simplified model in which a few action terms and fields are dropped and the part of the action that does not involve $\varphi$ is disregarded, not to speak of the quantum corrections. But all the ignored terms can be re-introduced in the analysis later on.

The equation of motion for $\varphi$ is 
\be
\ddot \Phi \Phi + \Phi\ddot \varphi +\dot\Phi \dot\Phi+\Phi \dot\Phi\dot\varphi=e^{-2\varphi} \left(\frac {6} {\kappa} \left( \frac {\ddot a}a + \frac {{\dot a}^2}{a^2} + \frac k{a^2}\right)+\frac 6{\kappa} \left(-\ddot\varphi+ \dot\varphi\dot\varphi \right)- m^2 \Phi^2  - \frac {2\mathfrak c}{\kappa} e^{-2\varphi}\right),\label{eom1}
\ee
while for $\Phi$ is
\be
\ddot\Phi  + \Phi\ddot \varphi +2\dot\Phi \dot\varphi+\Phi \dot\varphi\dot\varphi=  m^2 e^{-2\varphi} \Phi -\frac{\lambda}2 \Phi^3 \label{eom2}
\ee
Now let us make the ansatz
\be
\Phi(t)=\frac {\alpha}t, \quad\quad \varphi(t) = \ln \left(\beta t\right), \quad\quad a(t)= \gamma\, t\label{ansatz}
\ee
where $\alpha, \beta,\gamma$ are coefficients to be determined. 
We obtain a relation for these coefficients and the constants of the theory by inserting \eqref{ansatz} into \eqref{eom1}
\be
\frac {6}\kappa\left(3+ \frac k{\gamma^2}\right) - \frac {2\mathfrak c}{\kappa \beta^2}- 2 m^2 \alpha^2+\alpha^2 \beta^2 -\frac \lambda 2 \alpha^4 \beta^2=0\label{coeff1}
\ee
Another independent relation is obtained directly from \eqref{eom2}:
\be
2+ \alpha=\frac { 2 m^2\alpha}{\beta^2} - \lambda \alpha^3\label{coeff2}
\ee
Nothing changes if we replace in \eqref{ansatz} $t$ with $t-t_0$, with arbitrary $t_0$. A new solution  can be obtained by replacing $t$  with $b t$, where $b$ is any positive real number. This residual scale invariance is clearly inherited from the conformal invariance of \eqref{Strunc}. It implies that a physical meaning can be attached only to ratios of different values of $t$.

Further relations can be gotten by the variation of the metric. The eom is
\be
R_{\mu\nu}-\frac 12 g_{\mu\nu}  \left(R -{\mathfrak c} e^{-2\varphi}\right)= 2 \kappa e^{2\varphi} T_{\mu\nu}^{(m)} \label{eom3}
\ee
Let us write the em tensor in the form familiar in cosmology, i.e., that of a perfect fluid
\be
T_{\mu\nu}^{(m)} = (\rho + P ) u_\mu u_\nu -P g_{\mu\nu} \label{Tmunucosm}
\ee
where $\rho$ and $P$ are the energy density and pressure, respectively.
In the rest frame $u_\mu =(1,0,0,0)$, from the $tt$ component of \eqref{eom3} one gets
\be 
\rho = \frac 3{2 \kappa \beta^2}\left( 1+\frac k{\gamma^2} -\frac {\mathfrak c}{6\beta^2} \right)\frac 1{t^4}, \quad\quad {\rm i.e.} \quad\quad \rho \sim  \frac 1{a^4}\label{rho}
\ee
From the $\begin{matrix}i\\i\end{matrix}$ component one gets
\be
\frac {\ddot a}a +\frac {\kappa}3 (\rho + 3P) e^{2\varphi}+\frac {\mathfrak c}6 e^{-2\varphi}=0\label{eom4}
\ee
from which it follows that
\be 
P\sim\frac 1{a^4}\label{eomP}
\ee
 It must be remarked that the continuity equation is different from the familiar one in cosmology
\be
\dot \rho = -3 \frac {\dot a} a \left(\rho +3 P+\frac {\mathfrak c}{6\kappa} e^{-4\varphi}\right) - \left( 2 \rho -\frac{\mathfrak c}{2\kappa} e^{-4\varphi}\right) \dot \varphi \label{continuity}
\ee

The previous analysis can be straightforwardly extended to include in the action other fields and terms.

We shall refer to the time profile (\ref{ansatz},\ref{rho},\ref{eomP}) as the conformal regime.

\subsection{The problem of scales}

Let us recall that in a matter dominated universe $a\sim t^{\frac 23}$, in a radiation dominated one $a \sim t^{\frac 12}$, while in a dS geometry $a\sim e^{\Lambda t}$. Thus the conformal regime is different from the latter three. Now, in the big bang picture the universe crosses several regimes:  an initial explosion of pure energy without matter, followed at some time after perhaps $10^{31} sec$ by a period of inflation, during which $a$ expands exponentially like in dS geometry. This is followed by a period of reheating, when the inflaton energy is transferred to the creation of particles, giving rise to a radiation dominated regime. Favored by the cooling due the expansion this is  followed by a matter dominated regime where nucleosynthesis, elctroweak symmetry breaking, QCD phase transition, etc., take place in succession. The conformal regime is different from the above mentioned three, and due to its relevance, if any, to very high energy physics, it may be thought appropriate only for the very early stage of the evolution, just after the big bang and before inflation takes place (provided of course that this phase can be described by field theory).

Now let us remark that in $S^{(c)}$ the cosmological constant $\mathfrak c$ is multiplied by the factor $e^{-4\varphi}$. If time increases by a power of 10, for instance from $10^{-43} sec$ to $10^{-33} sec$ the factor $e^{-4\varphi}$ decreases by a factor $10^{40}$. That is, the effective cosmological constant evolves by 40 orders of magnitudes or even more while spanning equivalent configurations of the theory due to conformal symmetry. The important thing here is that, while the cosmological constant evolves by this huge amount, the fermion and gauge part of the action is unaffected by this change. This suggests a possible application.

There is a longstanding problem facing any approach which aspires to unify gravity with the SM, due to the relation between the cosmological constant and the energy of the vacuum; more precisely the vacuum energy density due to gravitation is represented by
\be
\rho_\Lambda = \frac {\mathfrak c}{2\kappa} \label{cosmdensity}
\ee
The observed value is
\be
|\rho_\Lambda^{(obs)}|\sim 2 \times 10^{-10}erg/cm^3\label{Obslambda}
\ee
The trouble is that when we put together in a unique theory, as we have done above, gravity and matter, the matter field theory comes with its own vacuum energy. The latter is always a divergent quantity and can be estimated only using different cutoffs, \cite{Gellmann68,shifman79,carroll}. If one uses the QCD scale one finds $\rho_{vac}^{QCD} \sim 1.6\times10^{36} erg/cm^3$. If one uses the electroweak scale one finds $\rho_{vac}^{EW} \sim 3\times 10^{47} erg/cm^3$. Finally if the scale of the cutoff is the Planck mass, one gets $\rho_{vac}^{Pl} \sim 2\times 10^{110} erg/cm^3$. In any case the gap with \eqref{Obslambda} is gigantic, and one is obliged to imagine another unknown entry in the above calculations to fill in the gap.

This sounds utterly unnatural and constitutes the so-called cosmological constant problem, see \cite{weinberg89,weinberg00,vilenkin01} and references therein. But if we look instead at $S^{(c)}$, \eqref{totalactionc}, the problem takes a different turn. First, as just pointed out,  the fermion and gauge parts of the action, as well as the Yukawa coupling and the quartic scalar couplings, are unaffected by Weyl transformations. On the contrary the other terms in $S^{(c)}$, and in particular the cosmological constant term may undergo drastic changes under a Weyl transformation.   If we choose, for instance, a `gauge' $\varphi=0$ we reproduce the just mentioned unnatural situation, but if we choose a sufficiently negative value for $\varphi$, for instance a `gauge' $\varphi\approx -25$ or a similar one, the effective cosmological constant takes on a value for which the gravitational vacuum energy may be comparable with the value of the vacuum energy of the theory, whatever it may be.  As pointed out above what really matters is the ratio between these two times. These two `gauges' correspond, according to our background solutions to different times along the cosmological evolution.

The previous gauge fixing is not simply a formal manipulation. It means that we can deal in the same theory with very different scales of energy. Now, the point is that we are able to quantize field theories only via a perturbative series. Therefore, for instance, the smallness of the measured cosmological constant disappears compared to the quantum corrections of the SM. Simply it does not make much sense to juxtapose matter and gravity (if the cosmological constant represents its vacuum energy) in the same quantum theory. However, the theory $\cal TW$ is conformal invariant. Therefore we can quantize it at the scale (i.e. the `gauge') where the perturbative approach makes sense, and transfer the quantized results (renormalization and unitarity) to the other scales, much as we do in quantum gauge theories where we do quantization in our favorite gauge and then we prove gauge fixing independence. In the present case therefore what we have to do next is to do quantization and show that conformal invariance is preserved.

But before turning to quantization, let us stress the particular flexibility of the just outlined idea. In a cosmological framework the gigantic jump from the electroweak scale to the tiny cosmological constant may not be the only one. We may need other intermediate scales. They can be inserted in the above scheme as follows. With reference to the beginning of section 4, in particular to eq.\eqref{Rscalar}, let us introduce two dilaton fields $\varphi_1$ and $\varphi_2$, which transform like $\varphi$ above, and pose
\be
\widetilde R_{12} = R + 6\left(\partial \!\cdot\! S -S\cdot S\right), \label{Rscalar12}
\ee
where
\be
S_\mu=\epsilon\, \partial_\mu \varphi_1 +(1-\epsilon)\, \partial_\mu\varphi_2\label{Smu}
\ee
and $\epsilon$ is a real number. 
Next consider, as an example, the action
\be
S_{12}^{(c)} &=&-\frac 1{2\kappa} \int d^4x \sqrt{g}\, e^{-2\varphi_1} \left(\widetilde R_{12} + {\mathfrak c}\, e^{-2\,\varphi_1}\right)\0\\
&&+ \int d^4 x \,\sqrt{g} \left[ g^{\mu\nu} \sfD^{(2)}_\mu \Phi^\dagger \sfD^{(2)}_\nu \Phi-e^{-2\varphi_2}  M^2 \Phi^\dagger \Phi -\frac {\lambda}4 \left(\Phi^\dagger \Phi\right)^2\right]\label{Hscalar+-}
\ee
for a complex scalar field $\Phi$, where the possible subscript $\pm$ has been ignored and $ \sfD^{(2)}_\mu=\partial_\mu + \partial_\mu\varphi_2$. 

$S_{12}^{(c)}$ is confomal invariant, and with a suitable choice of the two `gauges' for $\varphi_1$ and $\varphi_2$, we can prepare the theory in a reasonable form for quantization whatever the intermediate scale for $M^2$ is.

\section{The quantum costs}

As shown above it is relatively easy to transform a classical local field theory containing matter and gravity into a conformally invariant one. The price is cheap, it is enough to introduce a dilaton field and suitably manipulate with it the terms which are not  by themselves confomal invariant, \cite{codello12} (but we shall see below that there may be limitations). The main question is whether the invariance survives the quantization process. Quantization, at least perturbative quantization, requires that propagators and vertices be unambiguously defined. This is the case for $\cal TW$ because the theory has been constructed in order to guarantee this requirement, namely absence of O-type anomalies. The next sensible requirement for an effective theory, as was noted above, is unitarity. The icing on the cake would be the proof of renormalizabilty. In \cite{BG24} these issues were broached in the form of a brief review of the existing literature.

To illustrate them without facing the complex technicalities of a detailed treatment, I will quote one example. In \cite{oda22b} the author focuses on a theory defined by the classical action \eqref{scalaraction}.  As noted, this theory is Weyl-invariant and is known as Weyl-invariant scalar-tensor gravity. The author fixes both the diffeomorphsms and the Weyl gauge and works out the corresponding BRST symmetries. The quantization is carried out in the canonical way. The author computes the equal time (anti)commutation relations among all the fields and their conjugates. On this basis, he is able to prove the existence of a global symmetry the `choral' symmetry, which is spontaneously broken at the quantum level. Its Nambu-Goldstone bosons are the graviton and the dilaton, which are consequently massless.  The author is also able to analyse the physical S-matrix using the method introduced by Kugo and Ojima, \cite{kugo-ojima} and prove that it is unitary. 

This important result, further generalized with more general actions in subsequent papers of the same author and collaborators, meets obstacles when faced with renormalization. For renormalization theory, see for instance \cite{piguetsorella}, requires that we introduce in the action all the terms with the right dimensions and the same symmetry as the original ones in \eqref{scalaraction}.  This would mean adding to the latter also \eqref{Weyltensoraction} (or \eqref{Sc'W}).  This fact triggers the appearance of a new entry. We can get an idea of how this works as follows. If we limit ourselves to the lowest order kinetic operator for the graviton $h_{\mu\nu}= g_{\mu\nu}-\eta_{\mu\nu}$ in \eqref{scalaraction} we find, after a suitable gauge fixing, $\alpha \square$, where $\alpha$ is a constant with the dimension of a square mass. The addition of the term \eqref{Weyltensoraction} brings in the kinetic operator a quartic derivative, which in the simplest case can be represented as $\beta \square^2$, where $\beta$ is a dimensionless constant. Disregarding the tensorial factor the propagator is proportional to  the inverse of $\alpha \square + \beta \square^2$ , i.e. the inverse of $-\alpha p^2 + \beta p^4$, which can be written as follows
\be
\frac 1{-\alpha p^2 + \beta p^4}= \frac 1{p^2 (-\alpha +\beta p^2)} = -\frac 1{\alpha} \left( \frac 1{p^2}- \frac 1{p^2-\frac {\alpha}{\beta}}\right)\label{propghmunu}
\ee
This inevitably introduces a quadratic pole with negative residue, corresponding to a negative norm state, which is likely to violate unitarity.

The occurrence of physical ghosts in similar gravity or gravity plus matter theories has been confirmed in several papers, \cite{stelle1977,julve-tonin1978,sibold21,sibold23,oda23,sibold24,oda24,oda24b}. Based on these results one can reasonably expect that the theory $\cal TW$, defined by the action $S^{(c)}$,  eq.\eqref{totalactionc}, may present problems both for unitarity and renormalizability. The algebraic renormalization procedure requires the addition of the action term \eqref{Weyltensoraction}, which is likely to break unitarity. In an effective approach we must privilege unitarity, but this prevents renormalizability. We therefore take $S^{(c)}$ to be an effective field theory of SM plus gravity, effective in the sense that it is not UV complete, but forms a conformal unitary approximation of a still unknown UV complete theory. In the sequel we put on standby the issue of renormalizability and, while studying the effective action, we rather keep an eye on unitarity.

\subsection{Weyl invariance and Weyl anomalies}

Let us recall that $S^{(c)}$ beside the invariance under the gauge transformations already present in $S$ and under diffeomorphims, possesses conformal (or Weyl) invariance. These three invariances cannot be treated on the same footing, as attested, for instance, by the research of ref.\cite{oda24}: it is impossible to fix the gauges for all three symmetries in a consistent way, since the usual procedure does not produce compatible corresponding BRST symmetries. Only for two of them (or for two combinations thereof)  is this possible. It is obvious that the fundamental symmetries are the gauge symmetry and the diffeomorphism one. In order to proceed with quantization they have to be both gauge fixed with the associate introduction of relevant ghosts, and produce compatible BRST symmetries. This has been done, for instance, in \cite{BG24} for theories like $\cal T$ and $\cal TW$. As for the conformal symmetry, it is rather popular in the literature to introduce a corresponding gauge vector field $C_\mu$, i.e. to treat it like, for instance, the Abelian gauge symmetry in QED. Although this is legitimate, it is not necessary. The point of view expressed in \cite{BG24} is different: a new gauge field of conformal invariance is not necessary because a `gauge field' is already there, it is $\varphi$. For we have shown above that by its means we can fully implement Weyl symmetry. In geometric language this defines an integrable Weyl structure\footnote{In \cite{BG24} and in a previous version of this paper there was a careless statement concerning the existence of the dilaton propagator without gauge fixing. It is true that a naive dilaton propagator exists, but the interconnected (non-diagonalizable) kinetic operator of the graviton-dilaton system is not invertible without a further partial gauge fixing.}.

Quantizing a theory like $S^{(c)}$ means, as already pointed out, fixing the gauge both for the gauge groups and the diffeomorphisms, inserting them in the action, introducing the corresponding ghosts and writing down their actions. Then one proceeds by singling out the quadratic kinetic terms for all the fields, in order to identify the corresponding propagators, and, finally, by determining the interaction vertices. Once we have these tools we can start calculating the Feynman diagrams; to start with, the one-loop ones. Some of them will be UV divergent and we suppose that such divergences can be absorbed with a redefinitions of the fields and constants (couplings and masses). Motivated by the above references we suppose  that this one-loop renormalization can be carried out possibly at the cost of introducing physical ghosts through the term \eqref{Weyltensoraction}. We obtain in this way a one-loop effective action $W^{(1)}$ in terms of renormalized fields and constants, with the same form as the classical action but in terms of redefined fields and renormalized constants.  What we want to discuss next is the fate of conformal invariance.

Let us start from the classical definition of the e.m. tensor for free matter fields interacting with a background metric, which is
\be
 T_{\mu\nu}= \frac 2{\sqrt g}\frac {\delta S}{\delta g^{\mu\nu}}\label{Tmunucl}
\ee
$S$ being the classical action. If the latter is Weyl-invariant the e.m. tensor is traceless. This follows from the classical Ward identity
\be
{\delta_\omega S}= \int d^4\, \left(\frac {\delta S}{\delta g^{\mu\nu}} \delta_\omega g^{\mu\nu} +\sum_i \frac  {\delta S}{\delta f_i} \delta_\omega f_i\right)=0\label{completeWIcl}
\ee
where $f_i$ denotes generic matter fields. For infinitesimal $\omega$, $\delta_\omega g^{\mu\nu}= -2\omega g^{\mu\nu}, \delta_\omega f_i= -2 y_i \omega f_i$ (where $y_i$ is 0 for gauge fields, 1 for scalars and $\frac 32$ for fermions, etc.). If the matter fields are on shell (with the exception of the gauge fields), i.e.  $\frac  {\delta S}{\delta f_i} =0$, it follows that $T_{\mu\nu} g^{\mu\nu}=0$ due to the arbitrariness of $\omega$.

The problem we have to consider in the case of $S^{(c)}$ is however more complicated because the metric is dynamical and there are multiple interaction terms that couple the fields in various ways. Invariance of the classical action $S^{(c)}$ under Weyl transformations is given by $\delta_\omega S^{(c)}=0 $. The procedure is the same as for \eqref{completeWIcl} except that the metric is not anymore a spectator, but we have to differentiate also the EH action $S^{(c\pm)}_{EH}$. The equation we obtain is 
\be
  R + 2\,{\mathfrak c}\,e^{-2\varphi}=2 \kappa\, e^{2\varphi} \,T^{(m)}, \quad\quad\quad{\rm where}\quad\quad T^{(m)}= g^{\mu\nu} T^{(m)}_{\mu\nu}\label{Riccieqmod}
\ee
which is the trace of the eom of $g_{\mu\nu}$, see eq.\eqref{eom3}. Here $T_{\mu\nu}^{(m)}$ denotes the e.m. tensor of all the matter fields (including the dilaton) coupled to the metric. 

For simplicity of notation here we have dropped the $\pm$ suffix, since the rest of this section applies indifferently to both left and right sector. This oversimplification does not affect in an essential way what comes in the sequel.
How do we have to interpret the above equation \eqref{Riccieqmod}? Classically it is a part (the trace) of the equation of motion for $g_{\mu\nu}$. We understand it as an equation that (partly) identifies a background solution over which a quantization will be carried out. We limit ourselves to the subclass of such solutions where a nontrivial background configuration is present only for the metric and possibly for the dilaton and some scalar field, while all the other fields represent fluctuations about the null configuration. There can be nontrivial solutions of this type, \cite{parker2009}, but as an introductory approach we focus on the very simple case  where the metric is the flat Minkowski one and the background of $\varphi$ vanishes. For the same reason we choose $\mathfrak c_\pm=0$. Therefore at the background level eq.\eqref{Riccieqmod} reads: 0=0. 

Some comments on nontrivial background solutions can be found in Appendix.

Now let us come to the fluctuating fields. First we split the action $S^{(c)}= S^{(c)}_0+ S^{(c)}_{int}$ into its free and the interacting part. $S^{(c)}_0$ contains only the kinetic quadratic terms in each separate field. $S^{(c)}_{int}$ contains all the interaction vertices (which are infinite in number because the metric and the dilaton have dimension 0)\footnote{We assume that possible linear terms in $\varphi$ are renormalized to zero by tadpole diagrams, \cite{peskin}.} . Perturbative quantization is based on the propagators derived from $S^{(c)}_0$ and on the above mentioned vertices. The e.m. tensor of the matter fields is obtained by expanding $S^{(c)}$ in $h_{\mu\nu}$ and selecting the first order in this expansion (the 0-th order is the action of the matter fields coupled to the flat metric). The first order splits into various pieces in which the matter fields are generally entangled, but, if we restrict ourselves to the lowest (non-interacting) order, the entanglement disappears and we find a sum of distinct e.m. tensors, one for each species of matter fields including the dilaton. For instance, the zero-th order e.m. tensor obtained in this way for fermios is
\be
 T^{(f)}_{\mu\nu}= \frac i4 \overline {\psi} \gamma_\mu
{\stackrel{\leftrightarrow}{\partial_\nu}}\psi+(\mu\leftrightarrow\nu)- \eta_{\mu\nu}\frac i2  \overline \psi \gamma^\lambda {\stackrel{\leftrightarrow}{\partial_\lambda}}\psi, \label{Tmunufermion}
\ee
and for Abelian gauge fields, after fixing the Lorenz gauge, is
\be
T^{(g)}_{\mu\nu} =-\frac 1{g^2}\left(\partial_\mu A_\lambda \partial_\nu A^\lambda + \partial_\lambda A_\mu \partial^\lambda A_\nu- \frac 12 \eta_{\mu\nu} \partial_\lambda A_\rho \partial^\lambda A^\rho\right)\label{Tmunugauge}
\ee
For the other fields the expression may be less simple, see \cite{BG24}. 

As for the role of $R$, present in the $S^{(c)}_{EH}$ action, its contribution at this order is null because it is at least quadratic in $h_{\mu\nu}$. Thus at the classical level eq.\eqref{Riccieqmod} reduces to
\be
T^{(m)} =0 \label{Tm=0}
\ee
on shell. 
Since the various e.m. tensors are disentangled from one another and the equations of motion without interactions reduce to the free ones for each separate species, the e.m. for each species turns out to be tracelss on shell.

The same things can be repeated also for the one-loop renormalized theory for we assume that the difference consists only in renormalized fields and couplings, instead of the classical ones. But in the quantum case we have to introduce also the ghosts, both for gauge and diffeomorphism symmetry. Therefore in the one-loop quantum case in the right hand side of \eqref{Riccieqmod} and in \eqref{Tm=0}, we will have also the contribution from the ghosts, see \cite{BG24}.

To complete the quantization of $S^{(c)}$ we must verify the one-loop validity of \eqref{Tm=0}. This consists in the calculation of the trace of the e.m. tensor of each species separately, thus we can utilize  the results found for free fields. We know that the e.m. tensor trace in general does not vanish because of anomalies. Any trace anomaly can be written in the form
\be
{\cal A}_\omega[g,f] = \int d^4x \,\sqrt{g}\,\omega\, F[g,f] \label{traceanomaly}
 \ee
 where $g=\{g_{\mu\nu}\}$ is the metric, $\omega$ is the Weyl transformation parameter, $f$ denotes any other field and $F$ is a local function of $g$ and $f$.  For instance, the e.m. trace of matter fields contain in general terms where the density $F$ takes the form of the quadratic Weyl density
\be
\EW^2=R_{\mu\nu\lambda\rho} R^{\mu\nu\lambda\rho}-2 R_{\mu\nu}R^{\mu\nu} +\frac 13 R^2,\label{quadweyl}
\ee
the Gauss-Bonnet (or Euler) density,
\be
E=R_{\mu\nu\lambda\rho} R^{\mu\nu\lambda\rho}-4 R_{\mu\nu}R^{\mu\nu} +R^2,\label{gausbonnet}
\ee
and the Pontryagin density,
\be
P=\frac 12\left(\varepsilon^{\mu\nu\mu'\nu'}R_{\mu\nu\lambda\rho}R_{\mu'\nu'}{}^{\lambda\rho}\right). \label{pontryagin}
\ee 
Other possible anomalies have densities 
\be
T_e[V]= F_{\mu\nu}F^{\mu\nu}, \label{gaugeaction}
\ee
and
\be
T_o[V]=\varepsilon^{\mu\nu\lambda\rho}F_{\mu\nu}F_{\lambda\rho}.\label{chern}
\ee 
for an Abelian field $V_\mu$ with $F_{\mu\nu}= \partial_\mu V_\nu -\partial_\nu V_\mu$, as well as others, which are listed in \cite{BG24}. These are all anomalies that do not involve the dilaton field $\varphi$. But there are other possible trace anomalies which explicitly involve $\varphi$. For instance 
those with densities 
\be
\widetilde R^2\quad\quad {\rm and} \quad\quad   \widetilde  R_{\mu\nu} \widetilde R^{\mu\nu} 
\label{Rtildesquare}
\ee
are also consistent Weyl  cocycles. We do not include the density with squared tilded Riemann tensor, because  a suitable sum of the three would boil down to the quadratic Weyl tensor anomaly \eqref{quadweyl}. 

All the above anomalies appear with a definite coefficient in front, depending on the field species which are integrated over (but not on the regularization used).

We have to mention also other cocycles, the trivial ones, or coboundaries. An example is given by the cocycle with density $\square R$. It satisfies the consistency conditions, and does appear in many instances, but its coefficient depends on the regularization used to compute it. Thus this coefficient cannot have any physical meaning. There is an easy way to get rid of this anomaly by subtracting from the effective action a suitable local term. For the above case, in particular, this term can be chosen to be, for instance, the integral of $R^2$ with the appropriate coefficient. However, for a reason explained further on, this is not going to be a good counterterm.

Concerning the odd parity anomalies, we have constructed the theory $\cal T$ so as to get rid of them (and they are not modified by introducing $\varphi$). But for the even parity trace anomalies the story is different. Let us recall that they do not obstruct the existence of propagators, therefore they are not dangerous from the point of view of quantization. But the even parity trace anomalies have the same sign in both chiral sectors and the coefficients in front of them are so random that it is impossible to cancel them adding up different species, except perhaps in very exotic models. In order to ensure the survival of conformal invariance while preserving locality there remain the  Wess-Zumino terms.

\subsection{Wess-Zumino terms}

Assuming that $\omega$ is an anticommuting Abelian field, any anomaly \eqref{traceanomaly} must satisfy the consistency condition
 \be
 \delta_\omega {\cal A}_\omega =0\label{traceanomalyWZ}
 \ee
 which expresses simply the fact that two subsequent Weyl transformations made in opposite order yield the same result. This is in fact an integrability condition. It means that, with the help of an auxiliary field $\sigma$, which transform as $\delta_\omega \sigma=-\omega$, we can construct a local functional ${\cal W}_{WZ}[\sigma,g,f]$, such that
 \be
 \delta_\omega  {\cal W}_{WZ}[\sigma,g,f]= - {\cal A}_\omega[g,f] \label{WZtrace}
 \ee
This functional can be explicitly constructed 
\be
 {\cal W}_{WZ} [\sigma,g,f] = \int_0^1 dt\, \int d^4x\,\sqrt{g(t)} F[g(t),f(t)]\,\sigma \label{WWZ}
 \ee
 where 
  \be
 g_{\mu\nu} (t) = e^{2\sigma t} g_{\mu\nu},\quad\quad {\rm so \,\,\,that}\quad\quad  \delta_\omega g_{\mu\nu}(t)= 2(1-t)\,\omega\,g_{\mu\nu}(t), \label{interpolatingmetric}
 \ee
 and 
 \be
 f(t) = e^{-y\, t\, \sigma} f, \quad\quad \delta_\omega f(t) = -y(1-t)\omega f(t)\label{f(t)}
 \ee
 where $y=0$ for a gauge field, $y=1$ for a scalar field. 

For the field $\varphi$ we put $f(t)\equiv\varphi(t)= \varphi+\sigma t$, thus
\be
\delta_\omega \varphi(t) =\omega(1-t), \quad\quad \frac d{dt} \varphi(t)= \sigma\label{varphit}
\ee

It can be easily proved that it satisfies \eqref{WZtrace}. For instance,  for the anomaly with density \eqref{gaugeaction} the WZ term has a particularly simple form
\be
{\cal W}_{WZ}[\sigma, g,V]\sim \int d^4x \sqrt{g}\, \sigma \, F_{\mu\nu}F^{\mu\nu} \label{WZFF}
\ee

In conclusion at one-loop we have the possibility to  recover conformal invariance for $W^{(1)}$  by adding to the one-loop renormalized action a few suitable WZ terms while identifying $\sigma$ with $-\varphi$. This addition brings in the effective action new (renormalizable) interaction terms. Let us call the new effective action $W^{(c,1)}$. 

\subsection{On the use of WZ terms and `gauge fixing'}

WZ terms do not simply restore Weyl symmetry, they may be used also to secure unitarity. The renormalization program prescribes that at every order of quantization we add all the counterterms compatible with the underlying symmetry, in the present case gauge invariance, diffeomorphism and Weyl symmetry. The first two are the fundamental symmetries that are guaranteed via the BRST formalism and the Slavnov-Taylor identities. Weyl symmetry is preserved by the corresponding Ward identity we have discussed above. But the theory $S^{(c)}$ contains another symmetry, which we will need for quantization. Let us see it.

To contain the size of formulas let us limit ourselves to the action
\be
S_1^{(c)}=-\frac 1{2\kappa} \int d^4x \sqrt{g}\, e^{-2\varphi} \left(\widetilde R + {\mathfrak c}_\pm\, e^{-2\,\varphi}\right)+ \frac 12  \int d^4 x \,\sqrt{g} \left[ g^{\mu\nu} \partial_\mu \Phi \partial_\nu \Phi+ \frac 16 R \Phi^2 -e^{-2\varphi}  m^2 \Phi^2 -\frac {\lambda}4\Phi^4\right]\label{Strunc1}
\ee
with the possible addition of  \eqref{Weyltensoraction}, which contains the essential features for the following discussion. If we express the action in terms of 
\be
\widetilde g_{\mu\nu} =e^{-2\varphi} g_{\mu\nu}\quad\quad {\rm  and} \quad\quad\widetilde\Phi= e^{-\varphi} \Phi\label{fieldredef}
\ee
$S_C$ remains the same like the fermionic and gauge part of  $S^{(c)}$, while \eqref{Strunc} becomes
\be
S_1=-\frac 1{2\kappa} \int d^4x \sqrt{\widetilde g}\, \left(R + {\mathfrak c}_\pm\,\right)+ \frac 12  \int d^4 x \,\sqrt{\widetilde g} \left[\widetilde g^{\mu\nu} \partial_\mu\widetilde \Phi \partial_\nu \widetilde\Phi +\frac 16 R\widetilde \Phi^2 -  m^2 \widetilde\Phi^2 -\frac {\lambda}4\widetilde\Phi^4\right]\label{Strunc2}
\ee
where $R=R(\widetilde g)$. From the path integral point of view what we have done is a field redefinition, with a trivial Jacobian (i.e. a Jacobian that does not contain derivatives), after which we can integrate out $\varphi$ and make it disappear from the game. Thus one can say that $\varphi$ is a St$\ddot{\rm u}$ckelberg field. 

Returning to the issue of symmetry, let us notice that although the action $S_1$ is not Weyl invariant, it exhibits this symmetry (i.e. $\widetilde g \to e^{2\omega} \widetilde g$ and $\widetilde \Phi \to e^{-\omega} \widetilde \Phi$) in all the terms  (which include $S_C$ and all the fermionic and gauge terms) except the soft ones, i.e. those field monomials with dimension less than 4. Let  us call this partial symmetry after the redefinition \eqref{fieldredef} Weyl-reduced symmetry. It is an important symmetry because it limits the number of possible counterterms in the quantization process.
 
The counterterms allowed in the renormalization process are all local integrable terms with the right dimensions, invariant under the three symmetries + the Weyl-reduced one, thus they include in particular all the terms in the action $S^{(c)}$ (excluding the non-minimally coupled ones, i.e. $\zeta_h=0$). In particular among the action terms given by $S_C$, eq.\eqref{Weyltensoraction}, and
\be
S_{C1}=\frac 1{\eta_1} \int d^4x \sqrt{g}\, \widetilde R^2 \quad\quad{\rm and}\quad\quad
S_{C2}=\frac 1{\eta_2} \int d^4x \sqrt{g}\, \widetilde R_{\mu\nu} \widetilde R^{\mu\nu}, \label{Sc12}
\ee
only $S_C$ is allowed, because $S_{C1},S_{C2}$ do not satisfy the request after the redefinition \eqref{fieldredef}. Where this exclusion mechanism is consistent in renormalization theory has to be further investigated.

The problem with these two terms is that they contain quartic derivatives of the metric and the dilaton, that is they introduce physical ghosts in the theory with the annexed risks for unitarity. Therefore it is good news that they are excluded. The only counterterm that remains is $S_C$. It contains four derivatives of the metric, therefore it can give rise to the problem illustrated in example \eqref{propghmunu}. A WZ term may provide a way out.

We know that there is an anomaly with the same density as $S_C$, \eqref{quadweyl}.
But recall that we also have the corresponding WZ term. In this case the WZ term has the same form as the anomaly with $\omega$ replaced by $\sigma=-\varphi$. Inserting it in the first quantized action $W^{(1)}$ we restore conformal invariance at one loop and obtain,  say, the conformal invariant effective action  $W^{(c1)}$. However the latter contain physical ghosts due to the counterterms. This is true for a generic `gauge' of $\varphi$. But suppose that we choose a `gauge' by fixing $\varphi$ to a suitable constant value, so that the WZ terms exactly cancels the corresponding counterterm. In this `gauge' the physical ghosts disappear and a possible violation of unitarity at one loop is removed. Due to Weyl invariance  we can assume that if unitarity holds for this gauge it can be extended to all values of $\varphi$. Notice that this is similar to \cite{sibold21, sibold23,sibold24}, where the unwanted negative norm state appears only at the tree level, but not in higher loops, a rather mild and controllable violation of unitarity.  Remark that the necessary cancelation would be impossible if in addition there were also the counterterms \eqref{Sc12}, because we have only one gauge fixing at our disposal. 

As for the trivial anomaly with density $\square R$ we have already noticed that it can be canceled by a counterterm with density $R^2$, But this introduces again physical ghosts. Therefore, in this case, to cancel the anomaly it is more convenient to use the corresponding WZ term, which introduces in the theory only interaction terms.

In conclusion it seems to be possible to renormalize $S^{(c)}$, without non-minimal couplings, at one loop, while preserving Weyl invariance and avoiding physical ghosts. Whether this is possible at higher loop order is an open problem.

Let us now return to the issue of `gauge fixing', i.e. making a specific choice for $\varphi$ among the infinite many physically realizable ones. As noted before, in a unique theory we can describe radically different physical situations. We can study unitarity and renormalization for a specific choice of the `gauge' for $\varphi$. As long as conformal invariance is preserved the results extends to all configurations of the dilaton. The problem next is to understand why a specific configuration for $\varphi$, say $\varphi_0=const$\footnote{Due to the residual scale invariance we can always rescale $t$ so that $e^{\varphi_0}{\mathfrak c}$ corresponds to the measured cosmological constant.}, describes the physics of the universe in the present era. This is sometime called the second cosmological constant problem. The example of a ferromagnetic material in 2d can help intuition. The source of magnetization in such materials is the spin of the electrons in incomplete atomic shells, each electron carrying one unit of magnetic moment. Such spins can be imagined to be attached to lattice points and to interact with the neighboring ones in such a way that the state of lowest energy corresponds to all the spins being aligned.
At temperature T = 0 all the spins are aligned. When the temperature increases the thermal motion destroys this order, but not completely if the temperature is low enough; there remains patches where the spins are all aligned, with the result that a finite magnetization survives. As $T$ reaches the critical temperature $T_c$ and goes beyond it, order is completely destroyed and magnetization vanishes (disordered or paramagnetic phase). Of course if we reverse the procedure in the direction of lower temperatures, we are going to see larger and larger patches of oriented spin reappearing. The system is characterized by a correlation length $\xi$ that becomes infinite at $T=T_c$.  The correlation length is interpreted as the average size of the polarized spin patches; the fact that at the critical point this becomes infinite, means that we have patches of any size. Therefore the physical picture does not change, not only when we rescale the system rigidly, but also when we rescale it with a varying scale from point to point. But this is precisely conformal symmetry. 

We can imagine something similar happening in our $\cal TW$ theory. At very high energy we expect conformal invariance to hold. In this regime all the configurations of $\varphi$ are equivalent. When the energy density decreases patches with different solutions for $\varphi$ start to appear and consolidate. Each patch $\varphi$ has a precise value depending on the time it broke off from the rest. Time plays the role of spin direction in the above example. Within this picture our present universe is thought to live in one of these patches with $\varphi$ fixed for ever. This way of figuring out the evolution of the universe in our theory, at least in the very early stage, is a resignation to the anthropic principle. But, in the theory $\cal TW$, there does not seem to be a viable alternative. The above is a mechanism to `fix the gauge'  for conformal invariance. What are the alternatives? The breaking of conformal symmetry cannot be an explicit one, of course, if one wants this theory to represent a faithful approximation to a fundamental one, in which no explicit breaking is allowed.  It cannot be a spontaneous breaking either, because that would require a potential with a minimum. But in a conformal invariant theory also the potential must be conformal invariant and cannot have minima. Perhaps a different mechanism might  exist in a UV completion of $\cal TW$. 

 \section{Conclusion}
 
 This paper is a continuation of \cite{BG24}. The model, proposed there, that incorporates both SM and gravity, in a form that avoids all the type-O anomalies, has been presented here in a simplified  version including only one metric, instead of two. It preserves however the basic structure of two sectors, left and right, with mirror fermions and scalars, as well as $SU(3)$ and $U(1)$ gauge fields, while the $SU(2)$ gauge fields as well as the metric are in common. Two subjects have been developed. The first is an interpretation of the right sector as dark matter, a rather attractive and reasonable idea, but still at the level of research project. The second concerns Weyl symmetry and its possible connection with cosmology on the applicative side and with unitarity and renormalization on the theoretical field theory side. It has been shown that a background solution of the Weyl invariant theory exists that represents a regime different from the well known DeSitter, radiation dominated and matter dominated ones, a solution that may apply to the very early stages of the universe. This solution also suggests interesting applications to the cosmological constant problem. On the quantum field theory side the subject of Weyl symmetry and Weyl anomalies, already developed in \cite{BG24} has been reviewed and an application of the WZ terms has been illustrated to the problem of one-loop quantization of the model to show that it may be used to secure unitarity.
\vskip 0,6cm
{\bf Acknowledgements}. I woud like to thank the organizers of the worshop {\it What comes beyond the standard model} 2025, for inviting me to  give a talk which has motivated this paper. 
\vskip 0,6cm

\section{Appendix. Non-trivial background}

If we plug the background solution of section 4.1 in the lhs of eq.\eqref{Riccieqmod} we see that it does not vanish (although $R$ does). Of course the rhs cannot vanish either, which means that some of the energy-momentum matter traces cannot vanish due to the presence of some background value of the involved scalar fields. Therefore the discussion of section 5.1 has to be improved by allowing for the presence of a non-trivial classical background for the metric, the dilaton and possibly other scalar fields. 

What should one do in this and similar cases? The first thing is to expand the involved fields into a classical and quantum part
\be
g_{\mu\nu}= g_{0\mu\nu} +h_{\mu\nu}, \quad\quad \varphi= \varphi_0+ \chi, \quad\quad
\Phi=\Phi_0+\phi, \quad\quad {\rm etc.}\0
\ee
In the previous example $ g_{0\mu\nu}$ is the FLRW metric and $\varphi_0, \Phi_0$ are given by eq.\eqref{ansatz}. We assume that the classical background satisfies \eqref{Riccieqmod} and look for a quantum version thereof. The same solution must of course satisfy \eqref{eom3}. Then we contract the latter with the inverse background metric $g_0^{\mu\nu}$ and subtract from the result eq.\eqref{Riccieqmod}. In the rhs we obtain 
\be
2\kappa e^{2\varphi_0} \left(g_0^{\mu\nu}\langle\!\langle  T^{(m)} _{\mu\nu}\rangle\!\rangle -
\langle\!\langle g_0^{\mu\nu} T^{(m)}_{\mu\nu}\rangle\!\rangle\right)\label{traceTm}
\ee
where $\langle\!\langle\cdot\rangle\!\rangle$ represent the first order quantization.
In the corresponding lhs we can safely assume that $\langle\!\langle h_{\mu\nu}\rangle\!\rangle=0$, which is the same condition fulfilled in section 5.1 (i.e. absence of first order in $h_{\mu\nu}$), while the term $\langle\!\langle e^{2\varphi}  \rangle\!\rangle=0$ is subtracted away. Thus 
eq.\eqref{traceTm} denotes the violation of conformal invariance at the lowest order. The expression
\be
g_0^{\mu\nu}\langle\!\langle  T^{(m)} _{\mu\nu}\rangle\!\rangle -
\langle\!\langle g_0^{\mu\nu} T^{(m)}_{\mu\nu}\rangle\!\rangle\label{traceTm2}
\ee
reproduces the expression of the trace anomaly proposed by Duff, \cite{duff1994,duff2020}. The first order expressions of these anomalies may in general be different from those analysed before because they may explicitly contain the background fields. When computing perturbative anomalies we have to use the perturbative cohomology in order to verify whether they satisfy the consistency conditions. 

The WZ consistency condition $\delta_\omega \Delta_\omega=0$ for the cocycle $\Delta_\omega$ is split according to the decomposition
\be
\delta_\omega = \sum_{i=0}^\infty \delta_\omega^{(i)},\quad\quad\Delta_\omega= \sum_{i=1}^\infty \Delta_\omega^{(i)}\label{cocycles}
\ee
In particular
\be
 \delta^{(0)}_\omega \Delta^{(0)}_\omega=0,\quad\quad
\delta^{(0)}_\omega \Delta^{(1)}_\omega + \delta^{(1)}_\omega \Delta_\omega^{(0)}=0, \quad\quad\delta^{(1)}_\omega \Delta^{(1)}_\omega + \delta^{(0)}_\omega \Delta_\omega^{(2)}+\delta^{(2)}_\omega \Delta_\omega^{(0)}=0, \quad\quad \ldots \label{descentWZ}
\ee
The basic BRST transformations are
\be
&&\delta_\omega^{(0)}h_{\mu\nu} =  \omega g_{0\mu\nu},  \quad \delta_\omega^{(1)}h_{\mu\nu}=2\omega h_{\mu\nu} ,\quad \delta_\omega^{(2)}h_{\mu\nu}=0,\quad ...\label{seriesdeltah}\\
&&\delta_\omega^{(0)}\varphi=- \omega \varphi_0, \quad \delta_\omega^{(1)}\varphi=-\omega \chi, \quad \delta_\omega^{(2)}\phi=0,\quad\ldots\\
&&\delta_\omega^{(0)}\phi=- \omega \Phi_0, \quad \delta_\omega^{(1)}\phi=-\omega \phi, \quad \delta_\omega^{(2)}\phi=0,\quad...\label{seriesdeltaphi}
\ee
An explicit example can be found in Appendix D of \cite{BG24}.
\vskip `1 cm

\end{document}